\documentstyle[12pt,epsf,axodraw]{article}

\newcommand{\hep}[1]{{#1}}
\newcommand{\lsi}{\raise0.3ex\hbox{$<$\kern-0.75em\raise-1.1ex\hbox{$\sim$}}}
\newcommand{\gsi}{\raise0.3ex\hbox{$>$\kern-0.75em\raise-1.1ex\hbox{$\sim$}}}
\newcommand{\lsim}{\mathop{\lsi}}

\renewcommand{\vec}[1]{{\bf #1}}
\newcommand{\stern}{}
\newcommand{\sumint}[1]{\mathop{\int \!\!\!\!\!\!\!\! \sum_{\,\,\,\,\,\,#1}}}
\newcommand{\proj}{{\cal P}}
\newcommand{\mint}{\int\!}
\newcommand{\mudr}{\bar{\mu}}
\newcommand{\intv}[1]{\int\!\frac{d\Omega_{{\bf v}_{#1}}}{4\pi}}
\newcommand{\intvp}{\int\!\frac{d\Omega_{{\bf v}'}}{4\pi}}
\newcommand{\half}{\mbox{$\frac12$}}

\newcommand{\im}{{\rm Im}}
\newcommand{\re}{{\rm Re}}

\renewcommand{\(}{\left(}
\renewcommand{\)}{\right)}
\renewcommand{\[}{\left[}
\renewcommand{\]}{\right]}

\newcommand{\mmdebye}{m^2_{\rm D}}
\newcommand{\mdebye}{m_{\rm D}}
\newcommand{\tr}{{\rm t}}
\newcommand{\trace}{{\rm tr}}
\newcommand{\disc}{{\rm disc}}

\newcommand{\vE}{\vec{E}}

\newcommand{\vk}{\vec{k}}
\newcommand{\vp}{\vec{p}}
\newcommand{\vq}{\vec{q}}
\newcommand{\vv}{\vec{v}}
\newcommand{\mal}{\cdot}

\newcommand{\deltav}{\delta^{(S^2)}}

\newcommand{\mref}[1]{(\ref{#1})}
\newcommand{\mlabel}[1]{\label{#1}}

\newcommand{\mcite}[1]{Ref.\ \cite{#1}}
\newcommand{\eq}[1]{Eq.\ \mref{#1}}
\newcommand{\eqs}[1]{Eqs.\ \mref{#1}}

\begin{document}
 
\setlength{\baselineskip}{0.6cm}
\newcommand{\nn}{\nonumber}
\newcommand{\figysize}{16.0cm}
\newcommand{\figtopspace}{\vspace*{-1.5cm}}
\newcommand{\figbottomspace}{\vspace*{-5.0cm}}
  
\renewcommand{\theequation}{\thesection.\arabic{equation}}

\begin{titlepage}
\begin{flushright}
NBI-HE-99-04
\\
March 1999
\end{flushright}
\begin{centering}
\vfill

\setlength{\baselineskip}{0.9cm}
{\LARGE \bf 

Diagrammatic approach 
to soft non-Abelian dynamics 
at high temperature

}
\setlength{\baselineskip}{0.6cm}
\vspace{1cm}

Dietrich B\"odeker \footnote{e-mail: bodeker@nbi.dk}

\vspace{1cm} { \em 
The Niels Bohr Institute,
Blegdamsvej 17, DK-2100 Copenhagen \O, Denmark}

\vspace{2cm}
 
{\bf Abstract}

\vspace{0.5cm}

\end{centering}

\noindent
The dynamics of soft ($|\vec{p}|\sim g^2 T$) non-Abelian gauge fields
at finite temperature is non-perturbative. The effective theory for
the soft scale is determined by diagrams with external momenta
$p_0\lsim g^2 T$, $|\vec{p}|\sim g^2 T$ and loop momenta larger than
$g^2 T$. We consider the polarization tensor beyond the hard thermal
loop approximation, which accounts for loop momenta of order
$T$. There are higher loop diagrams, involving also the scale $gT$,
which are as important as the hard thermal loops. These higher loop
contributions are characteristic for non-Abelian gauge theories and
their calculation is simplified by using the hard thermal loop
effective theory. Remarkably, the effective one-loop polarization
tensor is found to be gauge fixing independent and transverse at
leading order in the gauge coupling $g$. The transversality indicates
that this approach leads to a gauge invariant effective theory.

\vspace{0.5cm}\noindent


\vspace{0.3cm}\noindent
 
\vfill \vfill
\noindent
 
\end{titlepage}
 
\section{Introduction}
Even at very high temperatures, when the running gauge coupling $g$ is
small, perturbation theory for non-Abelian gauge theories breaks down
at the spatial momentum scale $g^2 T$ \cite{linde,gross}. For static
quantities, like the free energy or correlation lengths, one can
integrate out the high momentum modes 
\footnote{For spatial vectors I
use the notation $k=|\vk|$.  4-momentum vectors are denoted by $K^\mu
= (k^0,\vk)$, and the metric has the signature $(+---)$
.}
$(p\gg g^2 T)$ in perturbation theory using dimensional reduction
\cite{farakos,braateneffective}. Then, at leading order, the
non-perturbative physics associated with the scale $g^2 T$ is
described by a 3-dimensional Euclidean gauge theory, which can be
easily treated in lattice simulations.

Dynamical quantities, which are sensitive to the scale $g^2 T$, are
more difficult to deal with. Since one has to consider non-zero real
frequencies, one cannot use dimensional reduction.  A prominent
example for such a quantity is the rate for electroweak baryon number
violation \footnote{More precisely, the Chern-Simons number diffusion
rate, which is proportional to the baryon number violation rate close
to thermal equilibrium \cite{rubakov}.}, which, at leading order, is
entirely due to gauge field modes with spatial momenta of order $g^2
T$ \cite{khlebnikov}.

Fortunately, even for non-zero $p_0$, one can use perturbation theory
to integrate out high momentum degrees of freedom to obtain an
effective theory for the soft field modes.  At leading logarithmic
order, this effective theory is described by a Langevin equation
\cite{letter}. It determines the characteristic frequency of the soft
field modes as
\begin{eqnarray}
	p_0\sim g^4 T \log(1/g),
\end{eqnarray}
and it allows for a lattice calculation of the baryon number violation
rate \cite{moore.log}.

The most direct approach to such an effective theory is to compute
diagrams for soft external momenta, while the loop momenta are
restricted to be larger than $g^2 T$.  At one-loop order, the dominant
contributions are the so-called hard thermal loops 
\cite{pisarski}-\cite{taylor:generating},
which are saturated by loop momenta of order $T$. They lead to the
Debye-screening of electric interactions on the length scale
$1/(gT)$. Therefore, only the magnetic degrees of freedom are
non-perturbative.

For a long time it was assumed, that the characteristic frequency of
the soft magnetic modes is of order $g^2 T$. It was realized in
\mcite{asy}, that hard thermal loops drastically change this
picture. They lead to a strong (Landau-) damping of the soft,
non-perturbative dynamics. Considering the hard thermal loop resummed
propagator, the characteristic frequency of the soft magnetic modes
can be estimated as $p_0\sim g^4 T$ \cite{asy}.
 
In this paper, we investigate the polarization tensor for soft
external momenta beyond the hard thermal loop approximation. There are
higher loop contributions, which are not suppressed relative to the
hard thermal loops. They consist of self energy insertions on the hard
loops and of ladder-type diagrams, and they can be economically
calculated using the hard thermal loop effective theory. This paper
describes the calculation of the effective one-loop polarization
tensor, the leading logarithmic result was presented in
\cite{regensburg}.  Summing the leading logarithmic contributions from
all $n$-loop diagrams by solving a Boltzmann equation for the soft
field modes, one obtains the effective theory of \mcite{letter}.
Recently, this effective theory has also been obtained in
\cite{asy2}, the Boltzmann equation was also obtained in
\cite{asy2}-\cite{valle}.  A recent detailed derivation of the
Boltzmann equation can be found in \cite{blaizot99}.

The paper is organized as follows: Sect.~\ref{sec:htl} briefly reviews
some properties of hard thermal loops. Sect.~\ref{sec:ladders}
discusses higher loop contributions to the polarization tensor, which
can be as large as the hard thermal loops. Sect.~\ref{sec:eff}
contains the computation of the one-loop polarization tensor within
the hard thermal loop effective theory in a covariant gauge.  In
Sect.~\ref{sec:higher} we discuss higher loop contributions and the
relation of the diagrammatic approach to the Boltzmann equation. The
results are summarized and discussed in Sect.~\ref{sec:summary}.
Appendix A contains the expression for the sum over Matsubara
frequencies which is used in the main text. Finally, it is shown in
Appendix B, that the results of Sect.~\ref{sec:eff} are also valid in
Coulomb-like gauges.

\section{Hard thermal loops}
\setcounter{equation}0 \mlabel{sec:htl} 
Hard thermal loops are one-loop contributions to $n$-point functions
for external momenta small compared to $T$. They are saturated by loop
momenta of order $T$. Furthermore, the dominant contribution is
obtained when one momentum in the loop is on shell, $Q^2 =0$.  For 
the 
remaining propagators containing $Q$, one can use a large energy, or
eikonal approximation
\begin{eqnarray}
        \frac{1}{(Q+P)^2} \simeq \frac{1}{2 q}\, \frac{\! 1}{v\mal P}
        \mlabel{lee} \,\,\, ,
\end{eqnarray}
where $P$ is some linear combination of the external momenta, and
$v^\mu = Q^\mu/q$. The spatial part of $v$ is the velocity of the 
hard particles in the rest frame of the plasma. Due
to \eq{lee}, the angular integration over $\vv$ and the integration
over $q$ factorize.

The hard thermal loop 2-point function, which is the same in Abelian
and non-Abelian theories, reads \cite{weldon}
\begin{eqnarray}
	\delta \Pi_{\mu\nu} (P) = 
	\mmdebye 
	\[
		- g_{\mu 0} g_{\nu 0} 
		+ p_0 \intv{} \frac{v_\mu v_\nu}{v\mal P}
	\]
	\mlabel{propagator.3.2} .
\end{eqnarray}
The $q$-integration is responsible for the factor $\mmdebye$, which is
the leading order Debye mass squared.  All fields coupling to the
gauge fields contribute to the hard thermal loops in the same
functional form.  In a SU($N$) gauge theory with $N_f$ Dirac fermions and
$N_s$ scalars in the fundamental representation one has $\mmdebye=
(1/3) (N + N_s/2 + N_f/2)g^2 T^2$. Finally,  the integral $\int d\Omega_\vv$ is
over the orientation of the unit vector $\vv$.  

Since \mref{propagator.3.2} is transverse, 
\begin{eqnarray}
	p^\mu\delta \Pi_{\mu\nu}(P) =0
	,
\end{eqnarray}
it can be written in terms of the 3-dimensionally (3-d)
transverse and longitudinal projectors \cite{weldon} \footnote{Note
that the projectors in Eqs.\ \mref{projector.t}, \mref{projector.l}
have a sign different from the ones in Ref.\ \cite{weldon}.}
\begin{eqnarray}
  \proj^{ij}_{\rm t}(\vec{p}) &=& \delta^{ij} - \frac{p^ip^j}{p^2}, 
  \quad
  \proj^{\mu 0}_{\rm t}(\vec{p}) = 0
	\mlabel{projector.t},
\\
  \proj^{\mu\nu}_{\ell}(P) &=& \frac{p^\mu p^\nu}{P^2} - g^{\mu\nu}
  - \proj^{\mu\nu}_{\rm t}(\vec{p})
  \mlabel{projector.l}.
\end{eqnarray}
such that
\begin{eqnarray}
        \delta \Pi^{\mu\nu} (P) =
	\proj_{\rm t}^{ \mu\nu}(\vec{p})
      \delta \Pi_\tr (P) +
    \proj^{\mu\nu}_{\ell} (P)
 \delta \Pi_\ell (P)  
.
 \mlabel{tensor}
\end{eqnarray}

For momenta $p_0, p\lsim gT$, the hard thermal loop
\mref{propagator.3.2} cannot be considered as a small correction to
the tree level kinetic term, and it has to be resummed. In a covariant
gauge with gauge fixing parameter $\xi$, the resummed propagator reads
\cite{weldon}
\begin{eqnarray}
   \stern\Delta^{\mu\nu}(P) = 
    \proj_{\rm t}^{ \mu\nu}(\vec{p})
    \stern\Delta_{\rm t} (P)  +
    \proj^{\mu\nu}_{\ell} (P)
    \stern\Delta_{\ell} (P)  
    + \xi \frac{p^\mu p^\nu}{P^4}
   .
  \mlabel{tree}
\end{eqnarray}
The 3-d transverse and longitudinal propagators
\begin{eqnarray}
        \stern\Delta_{ {\rm t}, \ell}(P) =     
          \frac{1}{-P^2 + \delta \Pi_{ {\rm t}, \ell}(P)}.
        \mlabel{propagator} 
\end{eqnarray}
have poles at $p_0^2 = \omega^2_{{\rm t}, \ell}(p) > p^2$,  and they have a cut
for $P^2 <0$. 

For static quantities, like the free energy or equal time correlation
functions, it is convenient to work in the imaginary time
formalism. Then $p_0$ is an imaginary Matsubara frequency, and for
$p\lsim gT$ the dominant contribution is given by the $p_0 = 0$
modes. In this case the 3-d transverse fields are unaffected by hard
thermal loops, $\delta \Pi_\tr(0,\vp) = 0$, and the only effect of
hard thermal loops is the Debye screening of the longitudinal fields
due to $\delta \Pi_\ell(0,\vp) = \mmdebye$.

For dynamical quantities $p_0$ has to be analytically continued
towards the real axis. Different ways of approaching the real axis
correspond to different time ordering prescriptions.  For
definiteness, in this paper we will consider the case that
\begin{eqnarray}
  p_0 = \re(p_0) +  i\epsilon
  \mlabel{p0},
\end{eqnarray}
which gives the retarded propagator.  

When $p_0$ is non-zero, also the magnetic, or 3-d transverse fields
are screened (``dynamical screening'').  For $p \sim g^2 T$ this has a
dramatic effect \cite{asy}, since $\stern\Delta_{\rm t} (P)$ is
smaller than the free propagator by two powers of $g$ when $p_0\sim
g^2 T$. Therefore the transverse gauge fields perform only small
fluctuations for this frequency scale. In order to obtain large
fluctuations \footnote{For an instructive discussion, see
\cite{asy}.}, which are necessary for non-perturbative processes like
electroweak baryon number violation, one has to consider the small
frequency limit $p_0\ll p$ , in which case
\begin{eqnarray}
        \delta \Pi_\tr (P) \simeq 
	-i \frac{\pi}{4} \mmdebye \frac{p_0}{p} \qquad (p_0\ll p)
        \mlabel{deltapilimit} 
	.
\end{eqnarray}
Then the hard thermal loop resummed propagator becomes
\begin{eqnarray}
         \stern\Delta_\tr(P) \simeq \frac{1}{p^2} 
        \frac{i\gamma_p}{p_0 + i \gamma_p}
	\qquad (p_0\ll p)
        \mlabel{propagator.5.4},
\end{eqnarray}
where $\gamma_p = 4p^3/(\pi\mmdebye)$. Since $\gamma_p$ is of order
$g^4 T$ when $p\sim g^2 T$, \eq{propagator.5.4} indicates that the
characteristic frequency for the soft transverse modes is $p_0\sim g^4
T$ \cite{asy}.  The higher loop contributions discussed in this paper
lead to a logarithmic correction to this estimate.

\section{Beyond hard thermal loops}
\setcounter{equation}0 \mlabel{sec:ladders}
It will now become clear that the hard thermal loop approximation is
not sufficient for obtaining the correct effective theory for the soft
modes.  Consider a hard thermal loop and imagine adding a self energy
insertion on an internal line, where the additional loop momentum $K$
is of order $gT$. This gives the diagram in Fig.\
\ref{fig:selfenergy}(a), in which the hard loop momentum $Q$ is on
shell, $Q^2 = 0$. Since the external momentum $P$ is soft, the
momentum $Q+P$ is almost on shell.

\begin{figure}[t]
 
\vspace*{-4.0cm}
 
\hspace{1cm}
\epsfysize=25cm
\centerline{
        \hspace{-3cm}
        \epsfysize=25cm\epsffile{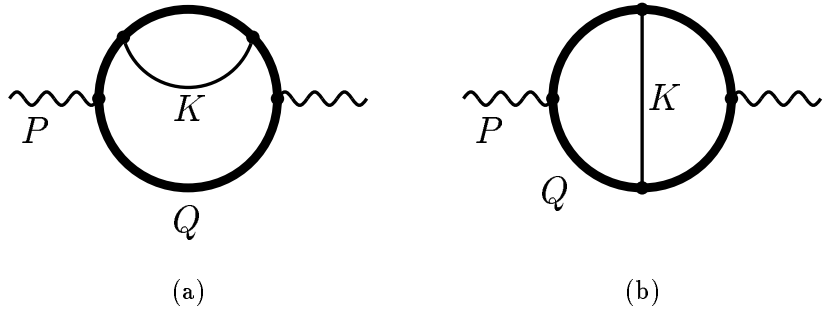}
        }

\vspace*{-16.5cm}
\caption[a]{Two-loop contribution to the polarization tensor.
  The external momentum $P$ is soft ($p_0$, $p \lsim g^2 T$) and the
  momentum $Q$ is hard ($q_0$, $q \sim T$). The thick lines denote
  propagators with momenta of order $T$. The thin lines are hard
  thermal loop resummed gauge field propagators $\stern \Delta(K)$
  with $k_0$, $k \sim g T$.  } \mlabel{fig:selfenergy}
\end{figure}
Compared to $\delta \Pi$, the diagram in
Fig.~\ref{fig:selfenergy}(a) contains two additional vertices, each of
order $g q\sim gT$. There are two additional propagators containing
the hard momentum $Q$, for which one can make the eikonal
approximation \mref{lee}, and there is one hard thermal loop resummed
propagator $\Delta(K)\sim 1/(g^2 T^2)$.  The loop integral over $K$
comes with a Bose distribution function
\begin{eqnarray}
        n(k_0) = \frac{1}{e^{ k_0/T} - 1}
        \mlabel{bose}
	,
\end{eqnarray}
which can be
approximated as  $n(k_0)\simeq T/k_0 $.
Thus we can estimate
\begin{eqnarray}
        \Pi^{\rm (1a)}(P) &\sim& 
         \delta \Pi (P) \times 
        \Bigg(g^2 T^2\Bigg) 
        \Bigg(\int\limits_{\,\,\, k\sim gT} \!\!\!  d^4 k \frac{T}{k_0}\Bigg)
        \Bigg(\frac{1}{T}\frac{\! 1}{v\mal P} \Bigg)
        \Bigg(\frac{1}{T}\frac{\! 1}{ v\mal (K+P)}\Bigg)
        \stern\Delta(K) 
        \nn\\
        &\sim& 
         \delta \Pi(P) \times 
        \frac{g^2 T}{v\mal P}
        \mlabel{pia}.
\end{eqnarray}
This shows that the diagram in Fig.\ \ref{fig:selfenergy}(a) can be as large as
$\delta \Pi(P)$,  when $p_0$ and $p$ are of order
$g^2 T$ or smaller.

Now consider the diagram in Fig.\ \ref{fig:selfenergy}(b). Proceeding as
above we estimate
\begin{eqnarray}
        \Pi^{\rm (1b)}(P) \sim 
         \delta \Pi(P) \times 
        \Bigg(g^2 T^2\Bigg) 
        \Bigg(\int\limits_{\,\,\, k\sim gT} \!\!\!  d^4 k \frac{T}{k_0}\Bigg)
        \Bigg(\frac{1}{T}\frac{\! 1}{v\mal K} \Bigg)
        \Bigg(\frac{1}{T}\frac{\! 1}{ v\mal (K+P)}\Bigg)
        \stern\Delta(K) 
        \mlabel{pib}.
\end{eqnarray}
On first sight, this diagram appears to be smaller than \mref{pia},
since one eikonal propagator $1/(v\mal P)$ got replaced by $1/(v\mal
K)$.  However, performing the integral over $k_0$, one obtains
contributions from the poles of the propagators. For example, at the
pole $k_0 = \vv\mal\vk$, the second propagator in \mref{pib} turns
into $1/(v\mal P)$.  Therefore we can estimate
\begin{eqnarray}
        \Pi^{\rm (1b)}(P) \! &\sim & \!
         \delta \Pi(P) \times 
        \Bigg(g^2 T^2\Bigg) 
        \Bigg(\int\limits_{\,\,\, k\sim gT} \!\!\! d^3 k \frac{1}{k}\Bigg)
        \Bigg(\frac{1}{T}\frac{\! 1}{v\mal P} \Bigg)
        \stern\Delta(K) 
        \nn\\ 
        \! &\sim&  \!
         \delta \Pi(P) \times  
        \frac{g^2 T}{v\mal P}
        \mlabel{pib2}\,\,\, ,
\end{eqnarray}
which is as large as \eq{pia}.

The above estimates for the diagrams in Fig.\ \ref{fig:selfenergy} are
well known and have been discussed extensively in the literature
\cite{lebedev,carrington1,carrington2}.  In order to deal with these
large contributions, the right half of diagram \ref{fig:selfenergy}(b),
together with the propagator $\stern \Delta(K)$ has been interpreted
as an insertion of a vertex correction. Then higher loop contributions
were summed using a Schwinger-Dyson equation. For Abelian theories it
was found that the large corrections cancel in the final answer.

It should be emphasized that this view is not convenient, because it
obscures the nature of the cancellation mechanism. Let us look at the
diagrams from a different perspective \cite{regensburg}. Imagine first
doing the integration over the hard momentum $Q$, with $K$ kept fixed.
This corresponds to integrating out hard momenta to obtain an
effective 4-point vertex for momenta small compared to $T$.  Now we
can make use of a well known fact: In an Abelian theory there are no
hard thermal loop vertices for gauge fields only.  In other words, the
large contributions due to the diagrams in Fig.~\ref{fig:selfenergy}
cancel in exactly the same way as the diagrams which, by power
counting, could give a hard thermal loop 4-point function.  In
particular, this means that the large contributions in an Abelian
theory cancel even at a given loop order, i.e., to see this
cancellation no resummation is necessary
\footnote{The cancellation at fixed loop order has already been
noticed in \mcite{carrington2}.}.

In contrast, in a non-Abelian theory, there are hard thermal loop $n$-point
functions for all $n$.  Thus in non-Abelian theories this cancellation
does not occur.

\begin{figure}[t]
 
\vspace*{-4.0cm}
 
\hspace{1cm}
\epsfysize=25cm
\centerline{\space{5cm}\epsffile{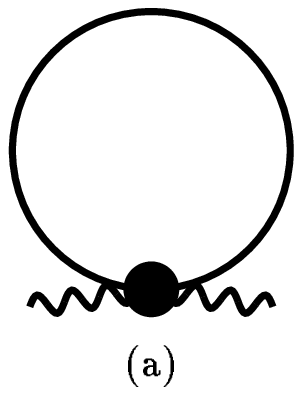}
        \hspace{-11cm}\epsfysize=25cm\epsffile{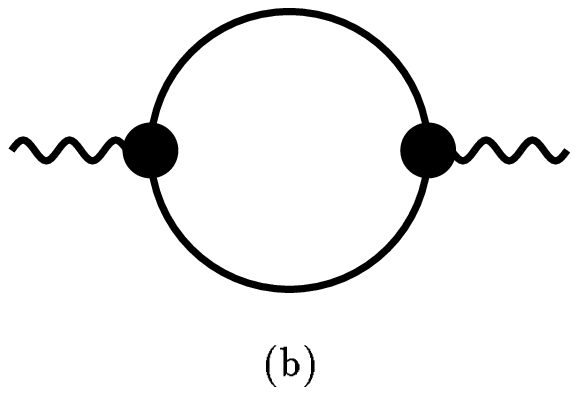}
        }

\vspace*{-18cm}
 
\caption[a]{One-loop  contributions to the polarization tensor 
  in the hard thermal loop effective theory. The heavy dots are
  hard thermal loop vertices. Otherwise the notation is the same as in
  Fig.\ \ref{fig:selfenergy}. }

\mlabel{htlvertex}
\end{figure}

\section{One-loop polarization tensor within the hard thermal loop
effective theory}\setcounter{equation}0
\mlabel{sec:eff} 
From the discussion in the previous section it should be clear that
the calculation of the diagrams in Fig.\ \ref{fig:selfenergy} become
simpler, if one uses the hard thermal loop effective theory.  Then
they correspond to a single
diagram, which is shown in Fig.\ \ref{htlvertex}(a).

The spatial loop momenta within the hard thermal loop effective
theory should be smaller than $\Lambda$, where 
\begin{eqnarray}
	gT \ll \Lambda \ll T.
\end{eqnarray}
We also introduce a scale $\mu$, which separates soft momenta from
momenta of order $gT$, such that
\begin{eqnarray}
	g^2 T \ll \mu \ll gT.
\end{eqnarray}
We consider only loop momenta larger than $\mu$, which yields an
effective theory for momenta smaller than $\mu$.

In addition to the propagator \mref{tree}, we will need the
expressions for hard thermal loop 3- and 4-point functions. They are
easily obtained from the hard thermal loop effective action in the
form given in Refs.\ \cite{blaizot93,Nair:local}. \footnote{For other
representations of the hard thermal loop effective action, see Refs.\
\cite{taylor:generating}, \cite{braaten.generating}-\cite{jackiw}.  A
recent and particularly simple derivation of the effective action of
Refs.\ \cite{braaten.generating,frenkel92} can be found in
\cite{elmfors}. For the relation of the different representations,
see, e.g., \cite{blaizot93,brandt95,pisarski97}.}  The $n$-point
vertex can be written as
\begin{eqnarray}
  \delta \Gamma^{a_1 \cdots
    a_n}_{\mu_1\cdots\mu_n}(P_1,\ldots,P_n) &=& g^{n-2} \mmdebye 
  \intv{} v_{\mu_1}\cdots v_{\mu_n} 
  \frac{1}{v\mal P_n}
  \nn\\ &&
  \hspace{-4.5cm}
  \Bigg\{2\, {\rm tr}(T^{a_n}[T^{a_{n-1}}, [\ldots ,T^{a_1}]\cdots ])
  \frac{p_1^0}{v\mal P_1} \frac{1}{v\mal(P_1 + P_2)}
  \cdots \frac{1}{v\mal(P_1 +\cdots + P_{n-2})} 
\nonumber\\
&&
  {}+ \mbox{permutations}[(P_1,a_1),\ldots,(P_{n-1},a_{n-1})]
  \Bigg\}
  \mlabel{deltaGamma} ,
\end{eqnarray}
where $T^a$ are the SU($N$) generators in the fundamental representation,
normalized such that ${\rm tr}(T^a T^b) = (1/2) \delta^{ab}$.

We use the imaginary time formalism \cite{bellac}. After
the sum over Matsubara frequencies has been performed, one can
analytically continue the external $p_0$ towards the real axis.

\subsection{Diagram 2a}
With the expression for the hard thermal loop 4-point function taken
from Eq.\ \mref{deltaGamma}, one finds
\begin{eqnarray}
        \Pi_{\mu\nu}^{(2 {\rm a})} (P) =
        \mmdebye N g^2 \intv{} 
        \frac{v_\mu v_\nu v_\rho v_\sigma }{(v\mal P)^2}\,\, 
        \sumint{K} \[ \frac{k_0+p_0}{v\mal (K+P)} 
        - \frac{k_0}{v\mal K}\] 
        \stern\Delta^{\rho\sigma}(K)
        \mlabel{4.1.4} ,
\end{eqnarray}
where 
\begin{eqnarray}
        \sumint{K} f(K)\equiv 
        T \sum_{k_0= i\omega_n} \mint \frac{d^3k}{(2\pi)^3} f(K) ,
\end{eqnarray}
and $\omega_n = 2\pi n T$ are the Matsubara frequencies with integer
$n$ running from $-\infty$ to $+\infty$.  

Performing the frequency sum can be quite tedious. One method, which
turns out to be quite convenient in the present case, is described in
Appendix A.  In order make use of Eq.\ \mref{sumd}, we rewrite the sum
appearing in \mref{4.1.4}
\begin{eqnarray}
        S&\equiv&
\sumint{K}
        \[ \frac{k_0+p_0}{v\mal (K+P)} - \frac{k_0}{v\mal K}\]
         \stern\Delta^{\rho\sigma}(K) 
         \mlabel{s3} ,
\end{eqnarray}
as 
\begin{eqnarray}
        S&=& -\frac12\,
        \sumint{K}
        \[ \frac{k_0+p_0}{v\mal (K+P)} - \frac{k_0}{v\mal K}\]
         \Big[ \stern\Delta^{\rho\sigma}(K +P) - \stern\Delta^{\rho\sigma}(K)
        \Big]
         \mlabel{s3a}.
\end{eqnarray}
Now we apply Eq.\ \mref{sumd} to obtain
\begin{eqnarray}
        S \!\!\! &\simeq& \!\!\! 
        -\frac12 T p_0\mint\!\frac{d^3k}{(2\pi)^3}
        \mlabel{tadpolesum}
\\ && \Bigg\{
        \frac{
        \stern\Delta^{\rho\sigma}(p_0,\vk + \vp) - 
        \stern\Delta^{\rho\sigma}(0,\vk)
        }{v\mal P - \vv\mal\vk}
        +
        \frac{
        \stern\Delta^{\rho\sigma}(0,\vk + \vp) - 
        \stern\Delta^{\rho\sigma}(-p_0,\vk)
        }{-p_0 - \vv\mal\vk}
\nn \\ &&{}
        - \mint \frac{dk_0}{2\pi i}
        \frac{
        \stern\Delta^{\rho\sigma}(K+\mbox{$\frac12$} P) - 
        \stern\Delta^{\rho\sigma}(K-\mbox{$\frac12$} P)
        }{(k_0 + \frac12 p_0)(k_0 - \frac12 p_0)}
        \Bigg[
        \frac{k_0 + \frac12 p_0}{v\mal (K+\frac12P)} - 
        \frac{k_0 - \frac12 p_0}{v\mal (K-\frac12 P)} 
        \Bigg] \Bigg\}
	\nn
	.
\end{eqnarray}
In this expression the analytic continuation of $p_0$ away from Matsubara
frequencies has already been performed. Furthermore, the high
temperature approximation for the Bose distribution function has been
used.

\eq{tadpolesum} looks quite complicated.  However, in the case, we are
interested in, which is $P\sim g^2 T$, $K\sim gT$, it can be
simplified considerably. Inside the integrand one can set $P=0$ except
for the imaginary part of $p_0$. The error made with this
approximation is of order $g$.
Then, the first two terms in the curly bracket
vanish, and one obtains
\begin{eqnarray}
	\mlabel{6.1.1}
        \Pi_{\mu\nu}^{(2 {\rm a})} (P) \simeq -\frac{i}{2}
         \mmdebye N g^2 T p_0 \intv{} \frac{v_\mu v_\nu
        v_\rho v_\sigma }{(v\mal P)^2} 
        \mint\!\frac{d^4k}{(2\pi)^4}\frac{1}{k_0}
        \disc\Big( \stern\Delta^{\rho\sigma}(K) \Big)
        \disc\(\frac{1}{v\mal K}\) 
        .
\end{eqnarray}
Eq.\ \mref{6.1.1} is gauge fixing independent in covariant
gauges. The terms proportional to the gauge fixing parameter $\xi$
contain factors $v\mal K$, which vanish due to 
\begin{eqnarray}
	\mlabel{disc}
	 \disc\(\frac{1}{v\mal K}\) =-i 2\pi \delta(v\mal K) .
\end{eqnarray}
Note that $v\mal K = 0$ is the approximate on-shell condition for the
loop momentum $Q+P$ in Fig.~\ref{fig:selfenergy}. In this case we have
\begin{eqnarray}
        (v\proj _{\rm t}(\vk) v) 
        = 
        - (v\proj _{\ell}(K) v) 
        =  1 - \frac{k_0^2}{k^2}  \qquad (v\mal K = 0)   ,
\end{eqnarray}
so that
\begin{eqnarray}
        \Pi_{\mu\nu}^{(2 {\rm a})} (P) &\simeq& -\frac{i}{2}
         \mmdebye N g^2 T p_0 \intv{} \frac{v_\mu v_\nu}{(v\mal P)^2} 
\nn \\ &&
        \mint\!\frac{d^4k}{(2\pi)^4}\frac{1}{k_0}
        \( 1 - \frac{k_0^2}{k^2} \)
	\disc\Big( \stern\Delta_{\rm t}(K) -\stern\Delta_{\ell}(K) \Big)
	        \disc\(\frac{1}{v\mal K}\) 
	\mlabel{tr.5.1}
        .
\end{eqnarray}
At this point one can see that $\Pi_{\mu\nu}^{(2 {\rm a})}$ is insensitive to
the UV cutoff $\Lambda$. Due to \eq{disc}, only space-like momenta
contribute to \mref{tr.5.1}. Then, the only discontinuity of the
propagators $\Delta_{{\rm t}, \ell}$ is due to the discontinuity of
the selfenergies, 
\begin{eqnarray}
	\mlabel{tr.3.3}
	\disc\Big(\Delta_{{\rm t}, \ell} (K) \Big) = 
	- |\Delta_{{\rm t}, \ell} (K)|^2 \,
	\disc \(\delta\Pi_{{\rm t}, \ell}(K) \) 
	\qquad ( K^2 < 0),
\end{eqnarray}
which falls off like $1/k^4$ for $k\gg\mdebye$.

However, the part containing the transverse propagator it sensitive to
the IR cutoff $\mu$, because $\Delta_{{\rm t}}(K)$ is unscreened for
$k_0\ll k$.  To compute the $\mu$-dependent piece of \mref{tr.5.1},
one can use the small frequency approximation \mref{propagator.5.4}
for $\Delta_{{\rm t}}(K)$ and neglect $k_0$ relative to $k$ otherwise.
Then the delta-function in \eq{disc} becomes $\delta(\vv\mal\vk)$, and
one obtains
\begin{eqnarray}
         \Pi_{\mu\nu}^{(2 {\rm a})} (P) \simeq -\frac{i}{4\pi}
         \mmdebye N g^2 T p_0 \intv{} \frac{v_\mu v_\nu
        }{(v\mal P)^2}\,\, \[\log\(\frac{gT}{\mu}\) +{\rm finite}\]
      \mlabel{pi4},
\end{eqnarray}
where ``finite'' denotes terms which are IR finite, i.e., which are
not sensitive to $\mu$. The finite terms will depend on how  the
cutoff is implemented.

\begin{figure}[t]
 
\vspace*{-4.0cm}
 
\hspace{1cm}
\epsfysize=25cm
\centerline{
        \hspace{-3cm}
        \epsfysize=25cm\epsffile{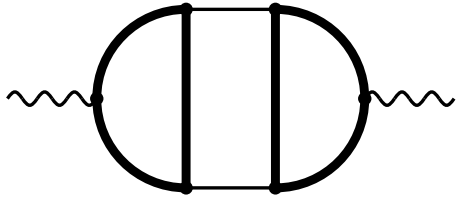}
        }

\vspace*{-18cm}
\caption[a]{Ladder-type diagram with two hard loop integrations
corresponding to the effective diagram in Fig.\ \ref{htlvertex}(b).
The notation is the same as in Fig.\ \ref{fig:selfenergy}.}
\mlabel{ladder}
\end{figure}

\subsection{Diagram 2b}\mlabel{sec:2b}
The result \mref{pi4} is gauge fixing independent, but it is not
transverse.  There is, however, another one-loop diagram in the hard
thermal loop effective theory, which contributes at the same order
(Fig.\ \ref{htlvertex}(b)). \footnote{The diagrams in Fig.\ 
  \ref{htlvertex} have been evaluated previously on the plasmon ``mass
  shell'' of the propagators \mref{propagator} to compute the gluon
  damping rate \cite{braatendamping} and corrections to the
  longitudinal plasmon frequency \cite{schulz}.  Our result does not
  apply to this case since we consider $p_0\ll gT$.}  In the
original theory it corresponds to a 3-loop diagram with two hard
loop momenta (Fig.\ \ref{ladder}).
The vertices can be read off from Eq.\ \mref{4.1.4}, and one obtains
\begin{eqnarray}
  \Pi_{\mu\nu}^{(2 {\rm b})} (P) &=& 
  -\half \mdebye^4 N g^2   
  \intv{1}\frac{v_{1 \mu} v_{1 \rho} v_{1 \sigma}}{v_1\mal P}
  \intv{2}\frac{v_{2 \nu} v_{2 \tau} v_{2 \lambda}}{v_2\mal P}
        \mlabel{1.6}
\\ && \hspace{-2cm}
  \,\,\sumint{K} 
  \stern\Delta^{\rho \tau} (K) \stern\Delta^{\sigma \lambda} (K+P)
  \[\frac{k_0}{v_1\mal K} - \frac{k_0 + p_0}{v_1\mal (K + P)}\]
  \[\frac{k_0}{v_2\mal K} - \frac{k_0 + p_0}{v_2\mal (K + P)}\]
  \nn .
\end{eqnarray}
Again, we use Eq.\ \mref{sumd} to perform the sum over Matsubara
frequencies, and the terms $f(0,p_0)$, $f(p_0,p_0)$ in \mref{sumd} do
not contribute at leading order in $g$. We neglect $P$ relative to $K$
except for the imaginary part of $p_0$ (see below), which gives
\begin{eqnarray}
  \Pi_{\mu\nu}^{(2 {\rm b})} (P) &\simeq& 
  -\frac{i}{2} \mdebye^4 N g^2 T  p_0
    \intv{1}\frac{v_{1 \mu} v_{1 \rho} v_{1 \sigma}}{v_1\mal P}
  \intv{2}\frac{v_{2 \nu} v_{2 \tau} v_{2 \lambda}}{v_2\mal P}
        \mlabel{10.2}
\\ && 
  \mint\!\frac{d^4k}{(2\pi)^4}
  \stern\Delta^{\rho \tau} (k_0 - i\epsilon, \vk) 
  \stern\Delta^{\sigma \lambda} (k_0 + i\epsilon, \vk) 
  \disc\(\frac{1}{v_1\mal K}\)
  \disc\(\frac{1}{v_2\mal K}\)
\nn .
\end{eqnarray}
Due to the delta-functions $\delta(v_i\mal K)$, only space-like $K$
contribute in \eq{10.2}.  Thus, there is no contribution from the
plasmon poles of the propagators at $k_0^2 = \omega^2_{{\rm t},
\ell}(k) > k^2$, which means that large denominators as in \eq{lee}
cannot arise here. Therefore, it was consistent to neglect $P$ in the
propagators.

Now we insert the expression for the propagator \mref{tree}. Again, the
terms proportional to the gauge fixing parameter $\xi$ drop out due to
the delta functions $\delta(v\mal K)$. Thus, we find that also the
leading order contribution from diagram 2(b) is gauge fixing
independent.  The remaining terms give
\begin{eqnarray}
  \Pi_{\mu\nu}^{(2 {\rm b})} (P) &\simeq& 
  -\frac{i}{2} \mdebye^4 N g^2 T  p_0
    \intv{1}\frac{v_{1 \mu} }{v_1\mal P}
  \intv{2}\frac{v_{2 \nu} }{v_2\mal P}
        \mlabel{tr.2.1.a}
\\ && 
  \mint\!\frac{d^4k}{(2\pi)^4}
        \Bigg| 
	\(v_{1} \proj_{\rm t} v_{2}\)
        \stern\Delta_\tr(K)
        + \(v_{1} \proj_{\ell} v_{2}\)
	\stern\Delta_\ell(K)
	\Bigg|^2
  \disc\(\frac{1}{v_1\mal K}\)
  \disc\(\frac{1}{v_2\mal K}\)
	\nn .
\end{eqnarray}
It is now easy to see that the $K$-integral is insensitive to
$\Lambda$: for $k\gg gT$ the propagators fall off like $1/k^2$.
Note that this integral also appears in the collision term in the
Boltzmann equation of Refs.~\cite{asy2,blaizot99}. There, the integrand
is given by the square of an on-shell matrix element.

Again, there is an infrared sensitive piece in the $K$-integral, which
is due to the term containing $ |\stern\Delta_{\tr}(K)|^2$. To compute
this piece, one can use the small frequency approximation
\mref{propagator.5.4} for $ \stern\Delta_{\tr}(K)$, and one can neglect
$k_0$ otherwise.  Using
\begin{eqnarray}
        \mint d\Omega_\vk  
	\(v_{1} \proj_{\rm t} v_{2}\)^2
	\delta(\vv_1\mal \vk ) \delta(\vv_2\mal \vk) &=&
        \frac{2}{k^2}
        \frac{(\vv_1\mal\vv_2)^2}{\sqrt{1 - (\vv_1\mal\vv_2)^2}} 
        \Theta\(-K^2\)
	,
\end{eqnarray}
one finds
\begin{eqnarray}
        \Pi_{\mu\nu}^{(2 {\rm b})} (P) &\simeq& 
        \frac{i}{\pi^2} \mdebye^2 N g^2 T p_0
        \intv{1}\frac{v_{1\mu}}{v_1\mal P} \intv{2} \frac{v_{2\nu}}{v_2\mal P}
        \nn\\ && {}
        \[ \log\(\frac{gT}{\mu}\) 
	\frac{(\vv_1\mal\vv_2)^2}{\sqrt{1 - (\vv_1\mal\vv_2)^2}} 
	+ {\rm finite} \]
        \mlabel{pi3}.
\end{eqnarray}

\subsection{Diagrams containing tree level vertices}
So far we have considered diagrams, which contain only hard thermal
loop vertices. Now we will see that diagrams
containing tree level vertices are smaller by one power of $g$.

The diagrams containing {\em only} tree level vertices have already
been discussed in Ref.~\cite{letter}. They are quite similar to hard
thermal loops since the loop momentum is large compared to the
external momentum. The size of hard thermal loops is determined by
integrals like
\begin{eqnarray}
	\int\limits_\Lambda^\infty d k k n(k) \sim T^2 
	\mlabel{htlintegral}.
\end{eqnarray}
In our case, $k$ is smaller than $\Lambda$, which means that one can
approximate $n(k)\simeq T/k$. Thus, instead of \mref{htlintegral}, we
have
\begin{eqnarray}
	\int\limits_\mu^\Lambda  dk k \frac{T}{k}.
\end{eqnarray}
The integration region $k\sim gT$
gives a contribution which is smaller than  the hard thermal loop
by one power of $g$. From $k$ near the cutoff $\Lambda$, one obtains a
contribution which cancels the $\Lambda$-dependence of the integral
\mref{htlintegral}~\footnote{The difficulties with this cancellation
which were encountered in Ref.~\cite{bms} are irrelevant as long as one
does perturbation theory.}.

\begin{figure}[t]
 
\vspace*{-3cm}
 
\hspace{1cm}
\epsfysize=25cm
\centerline{
        \hspace{2cm}
        \epsfysize=25cm\epsffile{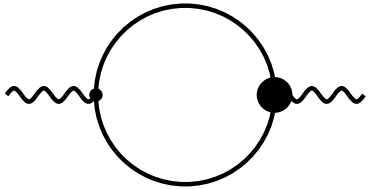}
        }

\vspace*{-19.5cm}
\caption[a]{Sub-leading contribution to the one-loop polarization
tensor with one hard thermal loop vertex and one tree level vertex.
The notation is the same as in Fig.\ \ref{htlvertex}.}
\mlabel{diagram.one}
\end{figure}

Now consider the diagram with one hard thermal loop and one tree level
vertex in Fig.\ \ref{diagram.one}, which gives
\begin{eqnarray}
  \mlabel{one}
  \Pi_{\mu\nu}^{(4)} (P) &=& 
  -\frac{i}{2} \mdebye^2 N g
  \intv{}\frac{v_{ \mu} v_{ \rho} v_{ \sigma}}{v\mal P}
\nn \\ 
&& \hspace{-2.5cm}
  \,\,  \sumint{K} 
  \stern\Delta^{\rho \tau} (K) \stern\Delta^{\sigma \lambda} (K+P)
  \Gamma_{\nu\lambda\tau}(-P, K+P, -K)
  \[\frac{k_0}{v\mal K} - \frac{k_0 + p_0}{v\mal (K + P)}\],
\end{eqnarray}
where $\Gamma$ is the tree level vertex.  After performing the
frequency sum, the square bracket gives a factor $\delta(v\mal K)$ in
the $P\to 0$ limit.  Thus, there are only contributions from space-like
$K$ and one can also take the limit $P\to 0$ in the propagators.
Since \mref{one} contains only a single factor $1/v\mal P$ rather
than two, it is suppressed relative to \mref{pi4}, \mref{pi3} by a
factor $g$.

\subsection{The logarithmic approximation}
Keeping only the terms in Eqs.\ \mref{pi4}, \mref{pi3}, which depend
logarithmically on $\mu$, the polarization tensor becomes \cite{regensburg}
\begin{eqnarray}
         \Pi_{\mu\nu}^{\rm( LA)} (P) = i \mdebye^2 N g^2 T 
         \log\(\frac{gT}{\mu} \)p_0
        \intv{1}\frac{v_{1\mu}}{v_1\mal P} \intv{2} \frac{v_{2\nu}}{v_2\mal P}
        I(\vv_1,\vv_2)
        \mlabel{pi34}.
\end{eqnarray}
Here, the function 
\begin{eqnarray}
  I(\vv_1,\vv_2) \equiv -\deltav(\vv_1 - \vv_2) 
        + \frac{1}{\pi^2}
        \frac{(\vv_1\mal\vv_2)^2}{\sqrt{1 - (\vv_1\mal\vv_2)^2}} 
  \mlabel{kern}
\end{eqnarray}
is the same as the one which appears in the noise correlator and in
the collision term in the Boltzmann equation of \mcite{letter} (cf.\
Sec.\ \ref{sec:higher}).  Furthermore, $\deltav$ is the delta-function
on the two dimensional unit sphere,
\begin{eqnarray}
        \int d\Omega_{\vv_1} f(\vv_1) \deltav(\vv - \vv_1) = f(\vv) .
\end{eqnarray}
One can easily verify that
\begin{eqnarray}
	\int d\Omega_{\vv_1} I(\vv_1,\vv_2) =0
	\mlabel{conservation.log} .
\end{eqnarray}
This implies that the polarization tensor \mref{pi34} is transverse
with respect to the external 4-momentum,
\begin{eqnarray}
	\mlabel{transverse}
        p^\mu \Pi_{\mu\nu}^{\rm( LA)} (P) = 0 .
\end{eqnarray}
This condition is necessary for the effective theory for the
soft fields to be gauge invariant.

\subsection{Beyond the logarithmic approximation}
In the previous section we found that the leading logarithmic result
for the polarization tensor is transverse. Now we will see that this
result holds beyond the leading logarithmic approximation, i.e., at
leading order in $g$. 

In order to define $\Pi_{\mu\nu}$ beyond the logarithmic
approximation, one has to specify the way how the loop momenta are cut
off in the infrared.  A convenient method is dimensional
regularization.  Then, the  above calculation of $\Pi_{\mu\nu}$ has to be
repeated in $n= 4-2\varepsilon$ instead of  4 dimensions, which is
straightforward.  In the hard thermal loop selfenergies and vertices,
the angular integration $\int d\Omega(\cdots)$ become $n-2$
dimensional, and in \eqs{propagator.3.2}, \mref{deltaGamma} one has to
replace
\begin{eqnarray}
	\int \frac{d\Omega_{\vv}}{4\pi}
	\to \int \frac{d\Omega_{\vv}}{\Omega} , 
\end{eqnarray}
where $ \Omega =\int d\Omega$.  Furthermore, one has to introduce a
scale $\mudr$ in order to keep the coupling constant dimensionless.

Then, the polarization tensor can be written as
\begin{eqnarray}
         \Pi_{\mu\nu}^{} (P) = i \mdebye^2 N g^2 \mudr^{2\varepsilon} T 
         p_0
        \int \frac{d\Omega_{\vv_1}}{\Omega}
	\frac{v_{1\mu}}{v_1\mal P} 
        \int \frac{d\Omega_{\vv_2}}{\Omega}
	\frac{v_{2\nu}}{v_2\mal P}
        \bar{I}(\vv_1,\vv_2)
        \mlabel{tr.6.3},
\end{eqnarray}
where 
\begin{eqnarray}
	\bar{I}(\vv_1,\vv_2) &=&  -\frac12
  \mint\!\frac{d^nk}{(2\pi)^n} 
	\mlabel{tr.7.1}
\\ &&
	\hspace{-1cm}
	\Bigg\{
		\Omega \delta^{(S^{n-2})}(\vv_1-\vv_2)
	\frac{1}{k_0} \(1 - \frac{k_0^2}{k^2} \) \disc\Big[
				\Delta_\tr(K) -\Delta_\ell(K)
			   \Big]
	  \disc\(\frac{1}{v_1\mal K}\)
\nn\\ &&
	\hspace{-1cm}
	{}+ \mmdebye
        \Bigg| 
	\(v_{1} \proj_{\rm t} v_{2}\)
        \stern\Delta_\tr(K)
        + \(v_{1} \proj_{\ell} v_{2}\)
	\stern\Delta_\ell(K)
	\Bigg|^2
  \disc\(\frac{1}{v_1\mal K}\)
  \disc\(\frac{1}{v_2\mal K}\)
	\Bigg\}
		\nn .
\end{eqnarray}
We will now see,  that $\bar{I}(\vv_1,\vv_2)$ also satisfies
\mref{conservation.log}, which then implies that $\Pi$ in \eq{tr.6.3}
is transverse.  We have
\begin{eqnarray}
        \int \frac{d\Omega_{\vv_1}}{\Omega}\bar{I}(\vv_1,\vv_2)
	&=& -\frac12
  \mint\!\frac{d^nk}{(2\pi)^n} 
	  \disc\(\frac{1}{v_2\mal K}\)
\nn \\ &&
	\hspace{-3cm}
	\Bigg\{
	\frac{1}{k_0} \(1 - \frac{k_0^2}{k^2} \) \disc\Big[
				\Delta_\tr(K) -\Delta_\ell(K)
			   \Big]
	\mlabel{tr.7.3}
\\ &&
	\hspace{-3cm}
	{}+ \mmdebye
	\int \frac{d\Omega_{\vv_1}}{\Omega}        \Bigg| 
	\(v_{1} \proj_{\rm t} v_{2}\)
        \stern\Delta_\tr(K)
        + \(v_{1} \proj_{\ell} v_{2}\)
	\stern\Delta_\ell(K)
	\Bigg|^2
  \disc\(\frac{1}{v_1\mal K}\)
	\Bigg\}
\nn  .
\end{eqnarray}
For the first term in the curly bracket one can use \eq{tr.3.3} to
write the discontinuities of the propagators in terms of the
discontinuities of the corresponding selfenergies. The latter can be
read off from (the $n$-dimensional version of) \eq{propagator.3.2}
making use of the projectors \mref{projector.t},
\mref{projector.l}. Taking into account that, in $n$ dimensions,
$\proj^{ii}_\tr = n - 2$, one finds
\begin{eqnarray}
	\disc\Big[
		\Delta_\tr(K) -\Delta_\ell(K)
	   \Big]
	&=&
	-\mmdebye k_0 \(1 - \frac{k_0^2}{k^2} \)
	\Bigg[
	\frac{1}{n-2}
        |\stern\Delta_\tr(K)|^2	
	+
	|\stern\Delta_\ell(K)|^2
	\Bigg]
\nn \\ &&
	\int \frac{d\Omega_{\vv_1}}{\Omega} 	
	\disc\(\frac{1}{v_1\mal K}\)
	\mlabel{tr.8.1}  ,
\end{eqnarray}
which contains precisely the same $\Omega_{\vv_1}$-integral as the second
term in the curly bracket of \eq{tr.7.3}.

In the second term in the curly bracket of \eq{tr.7.3} we can
perform the angular integration over $\vv_1$.  Due to $v_1\mal K =
v_2\mal K = 0$, the velocity vectors can be written as
\begin{eqnarray}
	\vv_i = \frac{k_0}{k}\hat{\vk} + \sqrt{ 1 - \frac{k_0^2}{k^2}}
	\, \hat{\bf \varphi}_i \qquad (i=1,2) ,
\end{eqnarray}
where $\hat{\bf \varphi}_i$ are unit vectors orthogonal to $\hat{\vk}
= \vk/k$. Then we have 
\begin{eqnarray}
	(v_{1} \proj_{\rm t} v_{2} ) & =&  \(1 - \frac{k_0^2}{k^2}\)
	\hat{\bf \varphi}_1 \mal \hat{\bf \varphi}_2 ,
\\
	(v_{1} \proj_{\ell} v_{2})& =&  -1 + \frac{k_0^2}{k^2}.
\end{eqnarray}
Now one can perform the $\hat{\bf \varphi}_1$-integration. The
transverse-longitudinal interference term in \eq{tr.7.3} vanishes due
to
\begin{eqnarray}
	\int d \hat{\bf \varphi}_1 \hat{\bf \varphi}_1 \mal \hat{\bf \varphi}_2
	=0.
\end{eqnarray}
Furthermore, we have
\begin{eqnarray}
	\int d \hat{\bf \varphi}_1 
	\(\hat{\bf \varphi}_1 \mal \hat{\bf \varphi}_2\)^2 
	= \frac{1}{n-2} \Omega .
\end{eqnarray}
Consequently, for $v_2\mal K = 0$, one can write 
\begin{eqnarray}
	\int \frac{d\Omega_{\vv_1}}{\Omega}        \Bigg| 
	\(v_{1} \proj_{\rm t} v_{2}\)
        \stern\Delta_\tr(K)
        + \(v_{1} \proj_{\ell} v_{2}\)
	\stern\Delta_\ell(K)
	\Bigg|^2
	\disc\(\frac{1}{v_1\mal K}\)
	= ~~~~~~~~~~~~~~&&
\nn \\
\mlabel{tr.8.2}  
	\(1 - \frac{k_0^2}{k^2} \)^2	
	\int \frac{d\Omega_{\vv_1}}{\Omega} 
	\Bigg[
	\frac{1}{n-2}
        |\stern\Delta_\tr(K)|^2	
	+
	|\stern\Delta_\ell(K)|^2
	\Bigg]	
	\disc\(\frac{1}{v_1\mal K}\) .
	&&	
\end{eqnarray}
Inserting \eqs{tr.8.1}, \mref{tr.8.2} into \mref{tr.7.3}, one finds that
the two terms in the curly bracket cancel, so that 
\begin{eqnarray}
	\int d\Omega_{\vv_1} \bar{I}(\vv_1,\vv_2) =0
	\mlabel{conservation.full}  .
\end{eqnarray}
This implies that $\Pi$ in \eq{tr.6.3} is transverse with respect to the
external $n$-momentum,
\begin{eqnarray}
	p^\mu   \Pi_{\mu\nu}(P) = 0.
\end{eqnarray}

\section{Higher loops}\setcounter{equation}0
\mlabel{sec:higher} 
In the previous section we have seen that there are one loop diagrams
which are as important for the soft field modes as the hard thermal
loop 2-point function. The natural question arises whether there are
higher loop diagrams which contribute at the same order and which  
have to be included in the effective theory for the soft fields as well.

The computation of higher loop diagrams along the lines of Sec.~4 can
be expected to be quite tedious. The number of terms in
the hard thermal loop $n$-point functions \mref{deltaGamma} grows
rapidly with $n$.  Moreover, performing the sum over Matsubara
frequencies in multi-loop diagrams is probably very difficult
(even for the one-loop diagram in Fig.~2b it was not particularly
easy).  A key simplification made in this paper was the high
temperature approximation for the Bose distribution function, which
corresponds to the classical field limit.  Unfortunately, this
simplification could be made only {\em after} performing the Matsubara
sum. Therefore it is more convenient to use an alternative formulation
of the hard thermal loop effective theory, which incorporates the
classical field approximation right from the start.

Fortunately,  such a formulation exists
\cite{blaizot93,Nair:local}. It is the non-Abelian generalization of
the linearized Vlasov equations for an electro-magnetic plas\-ma (see
e.g. \cite{landau10}). In addition to the classical gauge fields,
these equations contain fields $W^a(x,\vv)$ describing the
fluctuations of the phase space density of particles with momenta $\vq$
of order $T$ and $\vq/q = \vv$. The $W^a$ transform under the adjoint
representation of the gauge group.  In this formulation it is also
possible to integrate out momenta of order $gT$ perturbatively
\cite{letter, eff}. Again, one encounters terms which are
logarithmically sensitive to the se\-paration scale $\mu$. Keeping only
the leading logarithmic terms
one obtains an effective theory which is described by
the Boltzmann equation \cite{letter},
\begin{eqnarray}
        v \cdot D \,W(x,\vv) &=& \vec{v}\cdot\vec{E} (x)
        + \xi(x,\vv) -  \intvp C(\vv,\vv')      
 W(x,\vv') 
        \mlabel{boltzmann}
	,
\end{eqnarray}
together with the the Maxwell-Yang-Mills equation
\begin{eqnarray}
        \mlabel{maxwell}
        D_\mu F^{\mu\nu}(x)=  \mmdebye \int\frac{d\Omega_\vec{v}}{4\pi}
        v^\nu  W(x,\vv)
	.
\end{eqnarray}
$\xi$ is a Gaussian white noise which is specified by its 2-point function
\begin{eqnarray}
        \mlabel{xi0correlator4}
        \langle 
    \xi^{a}(x,\vv)
    \xi^{b}(x',\vv') 
    \rangle 
    =
        \delta^{ab} \delta^4(x - x')
      \frac{2 T}{\mmdebye} C(\vv,\vv')
	.
\end{eqnarray}
The integral kernel of the collision term reads
\begin{eqnarray}
	 C(\vv,\vv') = -  N g^2 T  \log\(\frac{g T}{\mu}\)I(\vv,\vv')
	, 
\end{eqnarray}
where $I(\vv,\vv')$ is given by \eq{conservation.log}.  Correlation
functions for soft external momenta are obtained by solving
\mref{boltzmann}, \mref{maxwell} and performing the thermal average
over initial conditions together with the noise average.

We will now make contact between the Boltzmann equation
\mref{boltzmann} and the one-loop calculation in
Sec.~\ref{sec:eff}. This will
demonstrate that there are indeed higher loop contributions to the
effective theory for soft field modes, which are summed by using the
Boltzmann equation.

Loosely speaking, the lhs of \eq{maxwell} is the first functional
derivative of the effective action which contains the effect of the
field modes with momenta larger than $\mu$.  Now we consider the
analogous quantity for the diagrammatic approach. We denote the
generating functional of the the ``tree level'' contribution
\mref{propagator.3.2} plus the one loop contribution \mref{pi34} by
$\widetilde{\Gamma}$.  The first (functional) derivative of
$\widetilde{\Gamma}$ is
\begin{eqnarray}
	\frac{\delta \widetilde{\Gamma}[A]}{\delta A^\nu(-P)} 
	=  \[ \delta \Pi_{\mu\nu} (P)
	+ \Pi_{\mu\nu}^{({\rm LA})}  (P) \]A^\nu (P)
	.
\end{eqnarray}
Obviously, this can be written in the form
\begin{eqnarray}
	 \frac{\delta \widetilde{\Gamma}[A]}{\delta A^\nu(-P)} 
	= \mmdebye \intv{} v^\nu \widetilde{W}(P,\vv)
	,
\end{eqnarray}
where
\begin{eqnarray}
	\widetilde{W}= \widetilde{W}_0+ \widetilde{W}_2
	.
\end{eqnarray}
Here the subscripts 0, 2 count the powers of $g$ which are not
contained in $\mmdebye$. $\widetilde{W}_0 $, corresponding to $\delta
\Pi$, reads
\begin{eqnarray}
	\widetilde{W}_0 = - A^0 + \frac{p^0}{v\mal P} \, v\mal A
	,
\end{eqnarray}
and $\widetilde{W}_2$, representing  $\Pi_{\mu\nu}^{({\rm LA})}$, is given by 
\begin{eqnarray}
	\mlabel{w2}
	\widetilde{W}_2 (\vv)= -i  \frac{p^0}{v\mal P}\intvp 
	\frac{1}{v'\mal P} \, C(\vv,\vv') v'\mal A
	.
\end{eqnarray}
\eq{w2} can also be written as
\begin{eqnarray}
	\widetilde{W}_2 (\vv)  = -i  \frac{1}{v\mal P}\intvp 
	C(\vv,\vv') \widetilde{W}_0(\vv')
	,
\end{eqnarray}
because due to \eq{conservation.log} the term $-A^0$ in
$\widetilde{W}_0$ drops out after integrating over $\vv'$.
$\widetilde{W}_0$ satisfies the usual linear Vlasov equation
\begin{eqnarray}
	\mlabel{boltzmann.w0}
	-i v\mal P \, \widetilde{W}_0 = 
	i v^i\[ - p^i A^0 + p^0 A^i \]  = (\vv\mal \vE)_{\rm linear}
	.
\end{eqnarray}
For $\widetilde{W}_2$ we have
\begin{eqnarray}
	\mlabel{boltzmann.w2}
	-i v\mal P  \, \widetilde{W}_2 (\vv) = - \intvp 
	C(\vv,\vv') \widetilde{W}_0(\vv')
	.
\end{eqnarray}
Adding \mref{boltzmann.w0} and  \mref{boltzmann.w2} we obtain  
\begin{eqnarray}
	\mlabel{boltzmann.w02}
	-i v\mal P  \, \widetilde{W}(\vv)
	=  (\vv\mal \vE)_{\rm linear} - \intvp 
	C(\vv,\vv') \widetilde{W}_0 ( \vv')
	.
\end{eqnarray}
Now we see that $\widetilde{W}$ satisfies the linearized Boltzmann
equation (without noise) up to terms which are higher order in the
loop expansion parameter $g^2$. Thus, from the validity of
\eq{boltzmann} one can infer that there are indeed higher loop
contributions to the leading order effective theory for the soft field
modes. Furthermore, the fact that the lhs of \mref{boltzmann} contains
a non-linear term implies that there are also vertex functions which
have to be included in the effective theory for the soft fields.

It is interesting to note that by writing the one loop result computed
in this paper as in \eq{boltzmann.w02}, one can already ``guess'' the
form of the Boltzmann equation. All one has to do to get from
\eq{boltzmann.w02} to \mref{boltzmann} is to replace $\widetilde{W}_0$
by $\widetilde{W}$ and to include non-linear terms to make the
equation gauge covariant. However, in this way one does not obtain a
noise term. A more quantitative discussion which accounts for the role
of the noise will be deferred to a separate publication.

\section{Summary and Discussion}
\setcounter{equation}0 \mlabel{sec:summary} 
We have seen that in non-Abelian gauge theories the hard thermal loop
approximation is not sufficient for obtaining the effective theory for
soft ($p\sim g^2 T$) momentum fields. When $p_0\lsim g^2 T$, the
polarization tensor for soft external momenta receives higher loop
contributions from loop momenta larger than $g^2T$, which are as large
as the hard thermal loops.

These large higher loop contributions are due to ladder type diagrams
and diagrams with selfenergy insertions on propagators carrying hard
momenta.  Their calculation is simplified if it is performed in two
steps. First, one integrates out the hard momenta, which gives the
well known hard thermal loop effective theory for momenta $p\ll
T$. Then, the polarization tensor is calculated within this effective
theory by integrating over momenta of order $gT$.
When the calculation is organized in this way, it is easy to see why
these large contributions cancel in an Abelian theory. The large
contributions discussed in this paper contain only hard thermal loop
gauge field vertices, which are absent in the Abelian case.

The calculation of the one-loop polarization tensor in the hard
thermal loop effective theory requires an explicit IR cutoff
separating spatial momenta of order $g^2T$ from momenta of order $gT$.
We have obtained an expression for the polarization tensor valid to
leading order in $g$, which is entirely due to diagrams containing
only hard thermal loop vertices. They correspond to a sum of 2- and
3-loop diagrams of the original theory. The result is gauge fixing
independent in covariant gauges and Coulomb-like gauges. Furthermore,
it is transverse with respect to the external 4-momentum, which is
necessary for the resulting effective theory to be gauge
invariant. The dependence on the cutoff is logarithmic, and the cutoff
dependent part was computed explicitly.

In the physically interesting region of very small  frequencies $p_0\ll
g^2 T$, only the 3-d transverse part of the polarization tensor is
unsuppressed relative to the hard thermal loop $\delta \Pi_\tr$. The
longitudinal part is small compared to $\delta \Pi_\ell$ , which, in the
small frequency limit,  approximately equals $\mmdebye$.

There are higher loop diagrams in the hard thermal loop effective
theory which are as important as the ones considered in this paper and which 
have to be included in the effective theory for the soft field modes.
Evaluating them using the present framework is quite tedious.  It is a
lot more convenient \cite{letter,eff} to use the formulation of the hard
thermal loop effective theory in terms of classical kinetic
equations. In this approach it is relatively easy to resum all leading
logarithmic contributions from higher loop diagrams.  One obtains a
Boltzmann equation for the field modes with $p<\mu$. 
The diagrams computed in this paper together with all leading higher
loop contributions are contained in a Gaussian white noise and a
collision term in the Boltzmann equation.~\footnote{In scalar field
theory the summation of ladder diagrams also leads to a Boltzmann
equation containing a collision term~\cite{jeon}.}

\vspace{1cm}

{\bf Acknowledgements.} I am grateful to Peter Arnold, Edmond Iancu,
Mikko Laine, Bert-Jan Nauta, and Toni Rebhan for useful discussions
and comments, and to Kimmo Kainulainen and Kari Rummukainen for
critical comments on the manuscript. This work was supported in part
by the TMR network ``Finite temperature phase transitions in particle
physics'', EU contract no. {ERBFMRXCT97-0122}.

\section*{Appendix A}\mlabel{appendix}
\renewcommand{\theequation}{A.\arabic{equation}}
\setcounter{equation}0
For diagrams containing the hard thermal loop resummed propagators, the
sum over Matsubara frequencies is not easy to compute.  The standard
method \cite{bellac} is to write the propagators in their spectral
representations. This method is not always useful. In particular, for
the diagram in Fig.~2b it turns out to be very inconvenient. This
Appendix describes an alternative method, which is very efficient when
applied to the diagrams considered in Sect.~4.

The sums are of the form
\begin{eqnarray}
        S = T \sum_{k_0 = i\omega_n} f(k_0, p_0) \mlabel{sum},
\end{eqnarray}
where $p_0 = i\omega_{n'}$ is a Matsubara frequency. For given $p_0$,
the function $f(k_0, p_0)$ has cuts in the complex $k_0$ plane for
$\im(k_0) = 0$ and for $\im(k_0) = - p_0$ (Fig.\ \ref{plane}). 

\begin{figure}
\begin{picture}(120,150)
        \Line(0,60)(120,60)
        \Line(60,0)(60,120)
        \multiput(60,12)(0,12){9}{\circle*{2}}
 \SetWidth{.1}
        \multiput(60,12)(0,12){9}{\ArrowArc(0,0)(4,0,360)}
 \SetWidth{2}
        \Line(0,60)(50,60)
        \Line(70,60)(120,60)
        \Line(0,24)(50,24)
        \Line(70,24)(120,24)
        
  \put(53,-20){(a)}
  \put(100,100){$k_0$}
  \put(50,130){Im($k_0$)}
  \put(127,57){Re($k_0$)}
 \SetWidth{.5}
        \Line(95,95)(95,110)
        \Line(95,95)(113,95)
\end{picture}
\begin{picture}(120,120)(-50,0)
        \Line(0,60)(120,60)
        \Line(60,0)(60,120)
        \multiput(60,12)(0,12){9}{\circle*{2}}
 \SetWidth{.1}
        \CArc(50,60)(4,270,90)
        \ArrowLine(0,64)(50,64)
        \ArrowLine(50,56)(0,56)
        \CArc(50,24)(4,270,90)
        \ArrowLine(0,28)(50,28)
        \ArrowLine(50,20)(0,20)

        \CArc(70,60)(4,90,270)
        \ArrowLine(70,64)(120,64)
        \ArrowLine(120,56)(70,56)
        \CArc(70,24)(4,90,270)
        \ArrowLine(70,28)(120,28)
        \ArrowLine(120,20)(70,20)
 \SetWidth{2}
        \Line(0,60)(50,60)
        \Line(70,60)(120,60)
        \Line(0,24)(50,24)
        \Line(70,24)(120,24)
        
  \put(53,-20){(b)}
\end{picture}
\begin{picture}(120,120)(-60,0)
        \Line(0,60)(120,60)
        \Line(60,0)(60,120)
        \put(60,34){\circle*{2}}
        \put(60,60){\circle*{2}}
        \put(60,34){\ArrowArc(0,0)(4,0,360)}
        \put(60,60){\ArrowArc(0,0)(4,0,360)}
 \SetWidth{.1}
        \ArrowLine(0,44)(60,44)
        \Line(60,44)(120,44)
        \Line(0,50)(60,50)
        \ArrowLine(120,50)(60,50)

        \ArrowLine(0,80)(60,80)
        \Line(60,80)(120,80)
        \Line(0,14)(60,14)
        \ArrowLine(120,14)(60,14)

 \SetWidth{2}
        \Line(0,60)(50,60)
        \Line(70,60)(120,60)
        \Line(0,34)(50,34)
        \Line(70,34)(120,34)
        
  \put(53,-20){(c)}
\end{picture}
\vspace*{2cm}
\caption[a]{Integration contours in the complex $k_0$ plane for the
integrals in (a) \eq{suma} and (b) \eq{sumb}. The contour in (c)
corresponds to the integral in \eq{sumd}. The lowest and highest
branches can be moved to $\pm i \infty$ and do not contribute. The
thick lines are the cuts of the function $f(k_0, p_0)$ located at
$\im(k_0) = 0$ and $\im(k_0) = -\im(p_0)$. } 

\mlabel{plane}
\end{figure}
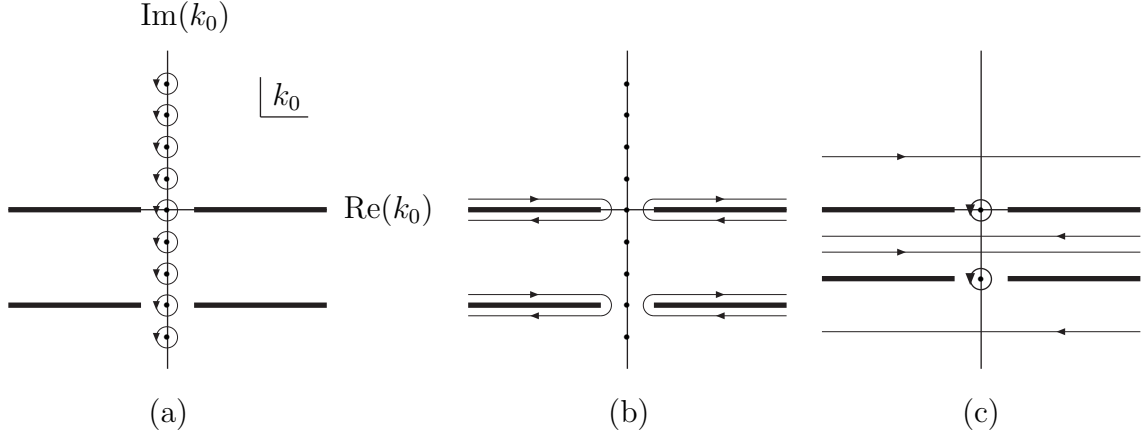
First we write the sum as an integral along the contour $C_{\rm a}$
depicted in Fig.\ \ref{plane}(a)
\begin{eqnarray}
        S
        = \int\limits_{C_{\rm a}}\frac{d k_0}{2\pi i}
        \(\frac12 + n(k_0)\)f(k_0, p_0)
        \mlabel{suma}
	,
\end{eqnarray}
where $n(k_0)$ is the Bose distribution function \mref{bose}.  Then
the contour is deformed as in Fig.\ \ref{plane}(b), which gives
\begin{eqnarray}
        S
        &=& \mint
        \frac{d k_0}{2\pi i}
        \(\frac12 + n(k_0)\)
        \Big(f(k_0+ i\delta, p_0) - f(k_0 - i\delta, p_0)\Big)
\nn\\ &&{}
        + \mint
        \frac{d k_0}{2\pi i}
        \(\frac12 + n(k_0)\)
        \Big(f(k_0 - p_0 + i\delta, p_0) - f(k_0 - p_0 
        - i\delta, p_0)\Big)
        \mlabel{sumb}.
\end{eqnarray}
Here the first (second) line corresponds to the upper (lower) part of
the contour. In the second line we have used the periodicity of the
function $n$, $n(k_0 + p_0) = n(k_0)$. Now we can perform the analytic
continuation of $p_0$ to arbitrary complex values. We are interested
in the case \mref{p0}. The $k_0$-integral is saturated for $|k_0|\ll
T$ so that we can use the high temperature approximation
\begin{eqnarray}
	\mlabel{classical}
	\frac12 + n(k_0)\simeq \frac{T}{k_0} 
\end{eqnarray}
to obtain
\begin{eqnarray}
        S &\simeq& 
        \mint
        \frac{d k_0}{2\pi i}
        \frac{T}{k_0}
        \Big(f(k_0+ i\delta, p_0) - f(k_0 - i\delta, p_0)\Big)
\nn\\ &&{}
        + \mint
        \frac{d k_0}{2\pi i}
        \frac{T}{k_0}
        \Big(f(k_0 - p_0 + i\delta, p_0) 
        - f(k_0 - p_0 - i\delta, p_0)\Big)
        \mlabel{sumc},
\end{eqnarray}
where $\delta < \im(p_0)$.
Now we deform the integration contour as in Fig.\
\ref{plane}(c): The upper part is closed around the pole at $k_0 =
0$. The piece above the cut can be moved towards $i\times\infty$
to give zero if $f(k_0, p_0)$ vanishes in this limit. The piece below the
the cut is moved downwards to Im$(k_0) = -{\rm Im}(p_0/2)$. Proceeding
similarly for the lower part of the contour, we finally obtain
\begin{eqnarray}
        S \simeq T\Bigg\{ f(0, p_0) + f(-p_0, p_0) 
        - p_0  \mint
        \frac{d k_0}{2\pi i}
        \frac{1}{k_0 + \half p_0} \frac{1}{k_0 - \half p_0}
        f(k_0 - \half p_0, p_0)\Bigg\}
        \mlabel{sumd}.
\end{eqnarray}

\section*{Appendix B}\mlabel{appendix.b}
\renewcommand{\theequation}{B.\arabic{equation}}
\setcounter{equation}0
In this Appendix it is shown that one obtains the same leading order
result for the polarization tensor as in Sect.~\ref{sec:eff},
if one uses  Coulomb gauge rather than a covariant gauge.

In Coulomb-like gauges the hard thermal loop resummed propagator reads
\begin{eqnarray}
	\Delta^{\mu\nu}  = \Delta_\tr \proj_\tr^{\mu\nu}
	+ \Delta_\ell \frac{K^2}{k^2} \, g^{\mu 0}g^{\nu 0}
	+ \xi \frac{k^\mu k^\nu}{k^4} .
	\mlabel{propagator.coulomb}
\end{eqnarray}
Strict Coulomb gauge corresponds to $\xi\to 0$. In order to see the
effect of using the propagator \mref{propagator.coulomb} instead of
\mref{tree}, it is convenient to write
$g^{\mu 0}g^{\nu 0}$ in terms of the tensors (cf. Ref.~\cite{gross})
$\proj_\ell^{\mu\nu}$,
\begin{eqnarray}
	C^{\mu\nu} = \frac{1}{\sqrt{2} k} \[
	\( 	g^{\mu 0} - \frac{k^\mu k^0}{K^2}  \) k^\nu
	+
	\( 	g^{\nu 0} - \frac{k^\nu k^0}{K^2}  \) k^\mu
	\],
\end{eqnarray}
and
\begin{eqnarray}
	D^{\mu\nu} =  -\frac{k^\mu k^\nu}{K^2}   .
\end{eqnarray}
Using 
\begin{eqnarray}
	\proj_\ell D = 0, \qquad \proj_\ell^2 = - \proj_\ell,
\end{eqnarray}
and
\begin{eqnarray}
	\trace(\proj_\ell) = \trace(D) = \trace(C^2) = -1, 
	\qquad \trace(C) = 	
	\trace(\proj_\ell C) =  \trace( C D ) =0,
\end{eqnarray}
one finds
\begin{eqnarray}
	\Delta^{\mu\nu}  = \Delta_\tr \proj_\tr^{\mu\nu}
	+ \Delta_\ell 
	\[
	\proj_\ell^{\mu\nu} + \sqrt{2} \, \frac{k^0}{k} C^{\mu\nu} 
	- \frac{k_0^2}{k^2} \, D^{\mu\nu} 
	\]
	+ \xi \frac{k^\mu k^\nu}{k^4} 
	\mlabel{propagator.coulomb.2}.
\end{eqnarray}
Now one can see that the propagator \mref{propagator.coulomb.2} gives
the same result as the propagator \mref{tree}. Each term in $C^{\mu\nu}$ and
$D^{\mu\nu}$ contains at least one factor $k^\mu$ or $k^\nu$, which
will be contracted with some velocity vector  $v$. Due to the 
delta functions $\delta(v\mal K)$ all these terms vanish.


\begin{thebibliography}{99}

\bibitem{linde} A.\ D.\ Linde, Phys.\ Lett.\ B 96 (1980) 289.

\bibitem{gross}
D.~J.~Gross, R.~D.~Pisarski and  L.~G.~Yaffe,
Rev.\ Mod.\ Phys.~53 (1981) 43.

\bibitem{farakos} K.~Farakos, K.~Kajantie, K.~Rummukainen and
  M.~Shaposhnikov, Nucl.\ Phys.\ B 425 (1994) 67,
hep-ph/9404201;
 Nucl.\ Phys.\ B 442
  (1995) 317,
hep-lat/9412091;
K.~Kajantie, M.~Laine, K.~Rummukainen and
  M.~Shaposhnikov, Nucl.~Phys.\ B 458 (1996) 90,
 hep-ph/9508379;
 A.~Jakovac,
  K.~Kajantie and A.~Patkos, Phys.~Rev.~D 49 (1994) 6810,
hep-ph/9312355.


\bibitem{braateneffective}  E.~Braaten and
  A.~Nieto, Phys.\ Rev.\ D 51 (1995) 6990,
hep-ph/9501375.

\bibitem{rubakov} For reviews of electroweak baryon number violation
and electroweak baryogenesis, see A.G.~Cohen, D.B.~Kaplan and
A.E.~Nelson, Ann.Rev.Nucl.Part.Sci.43 (1993) 27, hep-ph/9302210; V.A.\
Rubakov and M.\ E.\ Shaposhnikov, Usp.\ Fis.\ Nauk 166 (1996) 493,
hep-ph/9603208; A.~Riotto and M.~Trodden, CERN-TH-99-04, hep-ph/9901362.

\bibitem{khlebnikov} S.\ Yu.\ Khlebnikov, M.\ E.\ Shaposhnikov,
Nucl.\ Phys.\  B 308 (1988) 885.

\bibitem{letter}
D.~B\"odeker, Phys.~Lett. B 426 (1998) 351,
	\hep{hep-ph/9801430}.

\bibitem{moore.log}
G.D.~Moore, 
MCGILL-98-28, hep-ph/9810313.

\bibitem{pisarski}
E. Braaten and R. Pisarski, 
Nucl.\ Phys.\ B 337 (1990) 569;

\bibitem{frenkel90}
J. Frenkel and J.C. Taylor, 
Nucl.\ Phys.\ B 334 (1990) 199.

\bibitem{taylor:generating} J.C. Taylor and S.M.H. Wong, 
Nucl.\ Phys.\ B 346 (1990) 115.

\bibitem{asy}
P.\ Arnold, D.\ Son and L.G.\ Yaffe,
Phys.\ Rev.\ D 55 (1997) 6264,
\hep{hep-ph/9609481};
P.\ Huet and D.T.\ Son, 
Phys.\ Lett.\ B 393 (1997) 94,
\hep{hep-ph/9701393}.

\bibitem{regensburg} D.~B\"odeker, Proceedings of the {\em 5th International
              Workshop on thermal field theories and their
              applications}, ed. U.~Heinz, hep-ph/9811469.

\bibitem{asy2}
P.\ Arnold, D.\ Son and L.G.\ Yaffe,
UW/PT 98-10, MIT CTP-2779, hep-ph/9810216;
UW/PT 98-11, hep-ph/9901304.

\bibitem{litim} D.~Litim and C.~Manuel, ECM-UB-PF-99-04, CERN-TH-99-29,
hep-ph/9902430.

\bibitem{valle} M.A.~Valle Basagoiti, EHU-FT/9905, hep-ph/9903462.
 
\bibitem{blaizot99} J.\ P.\ Blaizot and  E.\ Iancu, Saclay-T99/026,
CERN-TH/99-71, hep-ph/9903389.

\bibitem{weldon}
See, e.g., H.\ A.\ Weldon, Phys.\ Rev.\  D 26 (1982) 1394.

\bibitem{lebedev} V.V.~Lebedev and A.V.~Smilga, Physica  A 181 (1992)
187.

\bibitem{carrington1}
M.E.~Carrington, R.~Kobes and E.~Petitgirard, 
Phys.\ Rev.\ D 57 (1998) 2631,
hep-ph/9708412;

\bibitem{carrington2}
M.E.~Carrington and R.~Kobes,
Phys.\ Rev.\ D 57 (1998) 6372,
 hep-ph/9712280.

\bibitem{blaizot93}
J.P.~Blaizot and E.~Iancu,
Phys.~Rev.~Lett.~{70} (1993) 3376,
hep-ph/9301236;
Nucl.Phys.\ B 417 (1994) 608, 
hep-ph/9306294.


\bibitem{Nair:local}
	V.P.~Nair,
	Phys.\ Rev.\ {D 48} (1993) 3432, 
	hep-ph/9307326.

\bibitem{braaten.generating} E. Braaten and R. Pisarski, Phys.\ Rev.\
D 45 (1992) 1827.

\bibitem{frenkel92}
J.~Frenkel and J.C.~Taylor,
Nucl.\ Phys.\ {B 374} (1992) 156.

\bibitem{efraty}
R.~Efraty and V.P.~Nair,
Phys. Rev. Lett.\ 68 (1992) 2891
hep-th/9201058; 
Phys.\ Rev.\ {D 47} (1993) 5601
hep-th/9212068.

\bibitem{jackiw} R.\ Jackiw and V.P. Nair, Phys.\ Rev.\ D 48 (1993) 4991
hep-ph/9305241.

\bibitem{elmfors}
P.~Elmfors, T.H.~Hansson and I.~Zahed,
Phys.\ Rev.\ {D 59} (1999) 045018,
hep-th/9809013.

\bibitem{brandt95}
F.T.~Brandt, J.~Frenkel and J.C.~Taylor,
Nucl.\ Phys.\ B 437 (1995) 433,
hep-th/9411130.

\bibitem{pisarski97}
R.D.~Pisarski,
hep-ph/9710370.

\bibitem{bellac}
For a review, see, e.g., 
M.~Le Bellac, {\it Thermal field theory}, Cambridge University Press (1997).

\bibitem{braatendamping} E.\ Braaten and R.\ Pisarski, Phys.\ Rev.\ D 42
  (1990) 2156.

\bibitem{schulz} H.\ Schulz, Nucl.\ Phys.\ B 413 (1994) 353.

\bibitem{bms} 
D.\ B\"odeker, L.\ McLerran and A.V.~Smilga, 
Phys.\ Rev.\ D 52 (1995) 4675,
\hep{hep-th/9504123}.

\bibitem{landau10}
E.M.~Lifshitz, L.P.~Pitaevskii, {\em Physical Kinetics} (Pergamon Press,
Oxford 1981).

\bibitem{eff}
D.~B\"odeker, NBI--HE--99--13, hep-ph/9905239, to appear in Nucl.\ Phys.\ B.


\bibitem{jeon} S.\ Jeon Phys.\ Rev.\ D 52 (1995) 3591; S.\ Jeon and 
L.\ G.\ Yaffe,
Phys.\ Rev.\ D 53 (1996) 5799.


\end{thebibliography}
\end{document}